\documentclass[prX,fleqn,showpacs,twocolumn]{revtex4}
\usepackage{latexsym}
\usepackage{amsfonts}
\usepackage{amsmath,revsymb}
\numberwithin{equation}{section}

\def\<{\langle}
\def\>{\rangle}
\def\+{\phi^+}
\def\-{\psi^-}

\def\b{\begin{equation}}
\def\bs{\boldsymbol}
\def\e{\end{equation}}

\def\G{{\cal G}}

\def\L{{\cal L}}

\def\s{{\mathsf{s}}}

\begin{document}

\title{A Bicycle Built for Two:  The Galilean and U(1) Gauge Invariance  of the Schr\"odinger Field.}
\author{V.~Colussi and S.~Wickramasekara}
\affiliation{Department of Physics\\
Grinnell College\\
Grinnell, IA 50112}

\begin{abstract}
This paper undertakes a study of the nature of the force associated with the local $U(1)$-gauge symmetry of a non-relativistic quantum particle. To ensure invariance under local $U(1)$ symmetry, a matter field must couple to a gauge field.  We show that such a gauge field satisfies Maxwell's equations, whether the matter field coupled to it is relativistic or non-relativistic. This result suggests that the structure of Maxwell's equations is determined by gauge symmetry rather than the symmetry transformation properties of space-time.  In order to assess the validity of this notion, we examine the transformation properties of the coupled matter and gauge fields under Galilean transformations. Our main technical result is the Galilean invariance of the full equations of motion of the $U(1)$ gauge field. 
\end{abstract}
\maketitle

\section{Introduction}\label{sec1}
	
As seen in textbook treatments  such as \cite{ryder}, 
electromagnetic fields and Maxwell's equations can be obtained from the requirement of the invariance of 
relativistic fields under local $U(1)$ gauge transformations. As the simplest case, consider 
a stable, spinless, relativistic particle of mass $m$. The Lagrangian density leading to the Klein-Gordon equation is 
\begin{equation}
\L=d_{\mu}\psi d^{\mu}\psi^*-m^2\psi\psi^*\label{1.1}
\end{equation}
This Lagrangian density is clearly invariant under global $U(1)$ transformations $\psi(x)\to e^{-i\lambda}\psi$ and $\psi^*\to e^{i\lambda}\psi^*$, where 
$\lambda$ is a real constant. Associated with this symmetry, by way of Noether's theorem, is the conservation of charge. In contrast, the Lagrangian density \eqref{1.1} is not invariant under local $U(1)$ gauge transformations 
\begin{equation}
\psi(x)\to e^{-i\lambda(x)}\psi(x),\quad \psi^*(x)\to e^{i\lambda(x)}\psi^*(x),\label{1.2}
\end{equation}
where $\lambda$ is now a real-valued function on the 3+1 dimensional Minkowski 	space. In order to obtain a local $U(1)$ gauge invariant Lagrangian density, it is necessary to couple the matter field of \eqref{1.1} to a gauge field, i.e., a four-component real valued function $A_\mu$ on the 3+1 dimensional Minkowski space that undergoes the transformations 
\begin{equation}
A_\mu(x)\to A_\mu(x)+\frac{1}{g}d_\mu\lambda(x)\label{1.3}
\end{equation}
when the matter field undergoes the transformations \eqref{1.2}. 
The Lagrangian density 
\begin{equation}
\L=(\partial_{\mu}\psi+ig{\bf{A}}_{\mu}\psi)(d^{\mu}\psi^*-ig{\bf{A}}^{\mu}\psi^*)-m^2\psi^*\psi-\frac{1}{4}F^{\mu \nu}F_{\mu \nu},
\label{1.4}
\end{equation}
where $F_{\mu\nu}=d_\mu A_\nu-d_\nu A_\mu$, remains invariant under \eqref{1.2} and \eqref{1.3}. The equations of motion for the gauge field that follow from \eqref{1.4} are precisely Maxwell's equations, where the charge-current densities are given in terms of the matter and gauge fields by $j^\mu=i\left(\psi^*(d^\mu+igA^\mu)\psi-\psi(d^\mu-igA^\mu)\psi^*\right)$. For a detailed derivation of these results, see for instance \cite{ryder}. Thus, we conclude that electromagnetic fields are what a complex matter field satisfying the Klein-Gordon equation must couple to in order to preserve invariance under local $U(1)$ gauge transformations. 

The foregoing discussion leads to the following natural question, the motivation of our study: what is the nature of the force associated with local $U(1)$ gauge invariance of a matter field satisfying the Schr\"odinger equation? There are two possibilities and each leads to further interesting questions. The first possibility is that this force is different from the electromagnetic force. If true, this would be quite perplexing since  there is ample empirical evidence that there are only four  fundamental forces in nature. Therefore, this possibility is quite unlikely. The second possibility is that the gauge force is still the electromagnetic force. If true, this possibility leads to a query about how Maxwell's equations, which provided the very motivation  for Einstein's special relativity, can arise from the gauge symmetry of non-relativistic field equations. The main conclusions of our study are that the $U(1)$ gauge invariance of the Schr\"odinger field in fact gives rise to Maxwell's equations and that these exact equations are invariant under Galilean transformations.

Now, it is widely acknowledged that Maxwell's equations are relativistic equations, and according to the standard historical account, it is the inconsistency of Maxwell's equations with Galilean relativity that led Einstein to his special theory of relativity. Textbooks present both electromagnetism and the structure of spacetime as a closed problem of sorts.  Nevertheless, there are several studies that examine the principles from which classical Maxwell's equations may be derived and their transformation properties under Galilean transformations. A significant, intriguing  contribution  is Dyson's paper on Feynman's proof of Maxwell's equations \cite{dyson}. Here, it is shown that a theory that assumes only Newton's laws of motion and the quantum mechanical commutation relations between position and velocity necessarily contains Maxwell's equations. In this regard, the conclusions of our paper are similar to the Dyson-Feynman result, albeit our starting point and approach involving $U(1)$ gauge symmetry are quite different from theirs.   In the conclusion of \cite{dyson}, Dyson writes ``here we find Galilean mechanics and Maxwell's equations coexisting peacefully''. In our view, this conclusion is not quite well supported in \cite{dyson} since Feynman and Dyson do not actually consider the transformation properties of their theory under Galilean transformations.  The explicit calculations presented here bear evidence to what Dyson saw as the ``peaceful coexistence" of Maxwell's equations and Galilean relativity.  

Following this rather tantalizing result of Feynman and Dyson, there have appeared a number of articles that investigate the origins and symmetry properties of Maxwell's equations. For instance, see \cite{dombey,brehme,anderson,horowitz,kapuscik}. An important article that precedes the Feynman-Dyson paper is that by Le Bellac and L\'evy-Leblond on the Galilean limit of Maxwell's equations \cite{bellac}. They arrive at the conclusion that  full Maxwell's equations are incompatible with Galilean relativity. This article is particularly important as it clearly sorts out many of the subtle issues associated with the Galilean limits of relevant quantities, both spacetime and electromagnetic. In fact, as pointed out in \cite{bellac},  many studies after Einstein have been dogged by a lack of rigour and the absence of a unified agreement on the characterization of what exactly is meant by a ``relativistic" effect and by a ``non-relativistic" effect, among other things.  Brown and Holland \cite{brown} also conclude the inconsistency of the full Maxwell's equations with Galilean relativity. On the other hand, Goldin and Shtelen \cite{goldin} assert that Maxwell's equations are consistent with Galilean relativity, but at the expense of nonlinear constitutive equations, i.e., a Galilean theory of electrodynamics is nonlinear.  Heras \cite{heras} takes a different approach and shows that Maxwell's equations can be obtained from the continuity equation for charge-current densities. Since the continuity equations is invariant under Galilean transformations, cf.~\eqref{3.3.20} below, here again we are prompted to ask whether or not Maxwell's equations respect Galilean relativity, despite their being able to be derived from assumptions that are consistent with Galilean relativity.  
 
Le Bellac and Levy-Leblond, as do many others, consider `Galilean electromagnetism' as a limiting case of the `exact theory', namely relativistic electromagnetism. They identify two different Galilean limits for electric and magnetic fields: the \emph{electric limit} where $|\bs{E}|>>c|\bs{B}|$ and the \emph{magnetic limit} where  $c|\bs{B}|>>|\bs{E}|$. As seen from \eqref{3.10} below and the discussion following it, corresponding to these two limits are two different Galilean limits of the Lorentz transformation formula for four-vectors, including spacetime vectors: the electric limit corresponds to $c|t|<<|\bs{x}|$ and the magnetic case to $c|t|>>|\bs{x}|$. Thus, starting with Maxwell's equations, one obtains two different electromagnetic theories in the Galilean limit. In the electric limit, electromagnetic fields and charge-current densities transform as
\begin{eqnarray}
	\rho_e^{\prime}&=&\rho_e\nonumber\\
	\bs{j}^{\prime}_e&=&\bs{j}_e-\bs{v}\rho_e\nonumber\\
	\bs{E}^{\prime}_e&=&\bs{E}_e\nonumber\\
	\bs{B}^{\prime}_e&=&\bs{B}_e-\frac{1}{c^2}\bs{v}\times\bs{E}_e\label{1.5}
\end{eqnarray}
while in the magnetic limit they transform as:
\begin{eqnarray}
	\rho^{\prime}_m&=&\rho_m-\frac{1}{c^2}\bs{v}\cdot\bs{j}_m\nonumber\\
	\bs{j}^{\prime}_m&=&\bs{j}\nonumber\\
	\bs{E}^{\prime}_m&=&\bs{E}_m+\bs{v}\times\bs{B}_m\nonumber\\
	\bs{B}^{\prime}_m&=&\bs{B}_m\label{1.6}
\end{eqnarray}
	
Notice that in neither case do we have full Maxwell's equations. In the electric case, a time-varying magnetic field does not induce an electric field, and, as a result, Faraday's law of induction is reduced to $\nabla\times\bs{E}_e=0$. In the magnetic limit, a time-varying electric field does not produce a magnetic field, leading to the requirement  $\nabla\cdot\bs{j}_m=0$.    

We wish to remark that our conclusion above about the Galilean invariance of Maxwell's equations does not quite contradict the result of Le Bellac and Levy-Leblond \cite{bellac} and others such as \cite{brown}.  The approach of both \cite{bellac} and \cite{brown} is to view Galilean transformations through the lens of special relativity, and if considered a `special case' of the relativistic theory, indeed one obtains the two sets of equations \eqref{1.5} and \eqref{1.6}. Our approach to the problem is fundamentally different in that we start with the Galilei group and push forward guided only by Galilean relativity and the principles of quantum physics. In particular, we take spacetime vectors to transform as $t'=t+b$ and $\bs{x}'=\bs{x}+\bs{v}t+\bs{a}$. As shown in section \ref{sec3} below, these transformation formulas imply that the gauge field be a Galilean vector field, transforming as in \eqref{3.2.4}. The only other result that we make use of is that the solutions to the Schr\"odinger equation for a particle of mass  $m$ and spin $s$ furnishes a unitary, irreducible, projective representation of the Galilei group, a well known result \cite{bargmann,levy-leblond}. These assumptions show that Maxwell's equations necessarily arise from the $U(1)$ gauge invariance of a matter field, be it relativistic or non-relativistic, and in the latter case, the full  Maxwell  equations are invariant under Galilean transformations. 

As a final remark, it should be pointed out that both the Feynman-Dyson result and our result are essentially quantum mechanical, whereas the Le Bellac and Levy-Leblond study is purely classical.  Neverthess, our techniques are different from that of Feynman and Dyson in that we take a Lagrangian approach, making use of the gauge  invariance of the Lagrangian density and equations of motion. While gauge symmetry is so central to modern theories of particle physics, which are relativistic, the implications of combined Galilean and gauge symmetry have not  been fully fleshed out.  Therefore, this paper can be seen as a first stride down a  parallel train-track of analogues between methods of relativistic quantum field theories and by comparison uncharted Galilean case. It is particularly interesting if the results of this paper can be meaningfully extended to non-Abelian gauge theories. 
	
The layout of the paper is as follows.  In section \ref{sec2}, we  introduce the Lagrangian density for the Schr\"odinger field and the resulting Euler-Lagrange equations, and we consider its invariance under $U(1)$ gauge transformations. In particular, we show that invariance under local $U(1)$ gauge transformations leads to Maxwell's equations. In Section \ref{sec3}, we study Galilean transformations of the $U(1)$ invariant Schr\"odinger Lagrangian density and equations of motion.

\section{Gauge symmetry of the Schr\"odinger field}\label{sec2}

Throughout this article, we assume a Galilean spacetime. However, it is convenient to consider $\mathbb{R}^4$ as our spacetime manifold and denote an element thereof by $x^\mu=(ct,\bs{x})$, $\mu=0,1,2,3$. Here, $c$ is a scaling constant with units of speed, but we need not attribute a physical meaning to it. We will also make use of notational conventions of special relativity such as the Einstein summation convention for repeated indices (be they upper or lower) and the use of Latin indices for spatial cooridnates, but clearly we do not also assume that the vectors $x^\mu$ of our spacetime transform as four-vectors of special relativity do. We adopt the notation 
\begin{equation}
d_\mu:=\frac{d}{dx^\mu}\label{2.1}
\end{equation}
The meaning of upper and lower indices will become clear in the next section when we discuss Galilean transformations.

The Lagrangian density for a free, non-relativistic quantum particle of mass $m$ and spin zero is 
\begin{equation}
\L=\frac{ic}{2}\psi^* d_0\psi-\frac{ic}{2}\psi d_0 \psi^*-\frac{1}{2m}\nabla\psi\cdot\nabla\psi^*\label{2.2}
\end{equation}
We have chosen units such that $\hbar=1$. The Euler-Lagrange equations for the fields $\psi^*(x^\mu)$ and $\psi(x^\mu)$,
\begin{subequations}
\label{2.3}
\begin{equation}
\tag{\ref{2.3}a}
d_\mu\frac{\partial \L}{\partial(d_\mu\psi^*)}\equiv d_0\frac{\partial\L}{\partial(d_0\psi^*)}+
\sum_{i=1}^{3}d_i\frac{\partial\L}{\partial(d_i\psi^*)}
=\frac{\partial\L}{\partial\psi^*}\label{2.3a}\\
\end{equation}
\begin{equation}
\tag{\ref{2.3}b}
d_\mu\frac{\partial \L}{\partial(d_\mu\psi)}\equiv d_0\frac{\partial\L}{\partial(d_0\psi)}+
\sum_{i=1}^3 d_i\frac{\partial\L}{\partial(d_i\psi)}
=\frac{\partial\L}{\partial\psi}\label{2.3b}
\end{equation}
\end{subequations}
immediately give the free particle Schr\"odinger equation for the wavefunction $\psi$ and its complex conjugate equation.  

The Lagrangian density \eqref{2.2} has several symmetries leading to, by way of Noether's theorem, conserved quantities. Recall that 
Noether's theorem states that if the action $I=\int_\Omega dx^\mu \L(\phi_l,d_\nu\phi_l,x^\mu)$ is invariant under transformations 
\begin{eqnarray} 
\Lambda: && \phi_l\to(\Lambda\phi)_l\nonumber\\
\chi: && x^\mu\to(\chi x)^\mu,\label{2.4}
\end{eqnarray}
where $\phi_l$ is a general $l$-component field, then there exists a conserved current 
\begin{equation}
d_\nu\left(\frac{\partial\L}{\partial(d_\nu\phi_l)}
\delta\phi_l+\delta x^\nu\L\right)=0\label{2.5}
\end{equation}
The variations $\delta\phi_l$ and $\delta x^\nu$ are defined as usual by 
$\delta\phi_l=(\Lambda\phi)_l-\phi_l$ and $\delta x^\nu=(\chi x)^\nu-x^\nu$.  

Often, as when the symmetry transformations constitute a finite dimensional Lie group, 
the transformation operators $\Lambda$ and $\chi$ are linear and can be defined 
as functions of a finite set of parameters $\epsilon_n,$ $n=1,2,3,\cdots,N$, and 
$\eta_m,$ $m=1,2,3,\cdots M$, respectively.
Further, as is again the case for Lie groups, let the values of $\epsilon_n$ and $\eta_m$ be such that
$\Lambda(\epsilon_n=0)=I$ and $\chi(\eta_m=0)=I$ and let there exist
neighborhoods $V$ and $W$ of the origins of the parameter spaces such that, to first oder, the operators
$\chi-I$ and $\Lambda-I$
depend on $\eta_m$ and $\epsilon_n$ {\em linearly} when
$\eta_m\in V$ and $\epsilon_n\in W$. For such $\eta_m\in V$ and
$\epsilon_n\in W$, we then
have
\begin{eqnarray}
\chi_{\mu\nu}&=&\delta_{\mu\nu}+\chi_{\mu\nu m}\eta_m\label{2.6}\\
\Lambda_{ll'}&=&\delta_{ll'}+\Lambda_{ll'n}\epsilon_n\label{2.7}
\end{eqnarray}
where $\chi_{\mu\nu m}$ and $\Lambda_{ll'n}$ are linear operators on
$x^\mu$ and $\phi$, and they may depend on the coordinates, field variables
and their derivatives but not on the parameters $\eta_m$ and
$\epsilon_n$. Substituting \eqref{2.6} and \eqref{2.7} into
\eqref{2.4}, we obtain to first order in $\epsilon$ and $\eta$, 
\begin{equation}
(\chi x)^\mu=x^\mu+(\chi_{\mu\nu m}x^\nu)\eta_m\equiv x^\mu+\overline{\chi}_{\mu m}\eta_m\label{2.8}
\end{equation}
\begin{eqnarray}
(\Lambda\phi)_l(x^\mu)
&=&\phi_l(x^\mu)-\chi_{\rho\nu m}x^\nu d_\rho\phi_l(x^\mu)\eta_m+
\Lambda_{ll'n}\phi_{l'}(x^\mu)\epsilon_n\nonumber\\
&\equiv&-\overline{\chi}_{\mu m}d_\mu\phi_l\eta_m+\overline{\Lambda}_{ln}\epsilon_n\label{2.9}
\end{eqnarray}
where $\overline{\chi}_{\nu
m}=\chi_{\nu\mu m}x^\mu$ and
$\overline{\Lambda}_{ln}=\Lambda_{ll'n}\phi_{l'}$.
The last equality of \eqref{2.9} makes use of the Taylor series
expansion of $\phi_l$ to first oder in $\eta_m$:
\begin{equation}
\phi_l(x^\mu-\chi_{\mu\nu m}x^\nu\eta_m)=\phi_l(x^\mu)-\chi_{\rho\nu
m}x^\nu d_\rho\phi_l(x^\mu)\eta_m\nonumber\\
\end{equation}
With \eqref{2.8} and \eqref{2.9}, the conserved Noether's current \eqref{2.5} acquires the form 
\begin{equation}
d_\nu\left(\frac{\partial\L}{\partial(d_\nu\phi_l)}
\overline{\Lambda}_{ln}\epsilon_n+
\overline{\chi}_{\mu m}\left(\delta_{\mu\nu}\L-\frac{\partial\L}
{\partial(d_\nu\phi_l)}d_\mu\phi_l\right)\eta_m\right)=0
\label{2.10}
\end{equation}
This equation tells us that there exists a conserved current 
associated with each independent symmetry parameter
$\epsilon_n$ and $\eta_m$. When $\eta_m$ are taken to be the parameters of Galilean transformations
and $\L$ is given by \eqref{2.2}, we obtain the conservation of energy, momentum, and angular momentum. We will discuss the Galilean 
transformations of \eqref{2.2} in greater detail in Section~\ref{sec3}.

\subsection{Global gauge transformations}\label{sec2.1}
For each $\lambda\in{\mathbb{R}}$, we define an operator $\Lambda(\lambda)$ on the two component Schr\"odinger field $\psi_l=\left(\begin{array}{c}\psi\\ \psi^*\end{array}\right)$ by
\begin{equation}
\Lambda(\lambda):\ 
\left(\begin{array}{c}
\psi\\
\psi^*
\end{array}\right)
\to 
\left(\begin{array}{c}
e^{-i\lambda}\psi\\
e^{i\lambda}\psi^*
\end{array}\right)\label{2.1.1}
\end{equation}
With respect to the usual inner product $(\psi,\phi)$, the operators $\Lambda(\lambda)$ are unitary, and the set $\{\Lambda(\lambda)\}$ is the commutative group $U(1)$ under composition, $\Lambda(\lambda_1)\Lambda(\lambda_2)=\Lambda(\lambda_1+\lambda_2)$. The transformations \eqref{2.1.1} are called global $U(1)$ gauge transformations. 

It follows readily that the Schr\"odinger Lagrangian density \eqref{2.2} is invariant under $\Lambda(\lambda)$. To determine the associated conserved current,  we note that $\overline{\chi}_{\nu m}=0$.  Then, letting $\overline{\Lambda}_{l=1}=\overline{\Lambda}_\psi$ and $\overline{\Lambda}_{l=2}=\overline{\Lambda}_{\psi^*}$, we obtain from \eqref{2.10}, 
\begin{equation}
d_\nu\left(\frac{\partial\L}{\partial(d_\nu\psi)}\overline{\Lambda}_\psi+\frac{\partial\L}{\partial(d_\nu\psi^*)}\overline{\Lambda}_{\psi^*}\right)=0\label{2.1.2}
\end{equation}
From the definitions \eqref{2.8}, \eqref{2.9}, and the transformation rule \eqref{2.1.1}, we have 
\begin{eqnarray}
\overline{\Lambda}_\psi&=&i\psi\nonumber\\
\overline{\Lambda}_{\psi^*}&=&-i\psi^*\label{2.1.3}
\end{eqnarray}
Substituting \eqref{2.2} and \eqref{2.1.3} in \eqref{2.1.2} gives 
\begin{equation}
d_0(c\psi^*\psi)+\nabla\cdot\left(\frac{i}{2m}\left(\psi\nabla\psi^*-\psi^*\nabla\psi\right)\right)=0\label{2.1.4}
\end{equation}
This continuity equation, along with the assumption that the fields $\psi$ and $\psi^*$ decay at infinity, gives the conservation of the global quantity $\int_{{\mathbb{R}}^3} d\bs{x}\psi^*(x^\mu)\psi(x^\mu)$, interpreted as total probability. Hence, we see that the conservation of probability of non-relativistic quantum physics is intimately connected with the invariance of the Lagrangian density --thus also of the action-- under global $U(1)$ gauge transformations.

\subsection{Local gauge transformations}\label{sec2.2} 

As an immediate generalization of the global gauge group $U(1)$, we now let the group parameter 
$\lambda$ be arbitrary, differentiable, real valued functions of $x^\mu$. That is,  
\begin{eqnarray}
(\Lambda(\lambda)\psi)(x^\mu)&=&e^{-i\lambda(x^\mu)}\psi(x^\mu)\nonumber\\
(\Lambda(\lambda)\psi^*)(x^\mu)&=&e^{i\lambda(x^\mu)}\psi^*(x^\mu)\label{2.2.1}
\end{eqnarray}
These operators are also unitary and fulfill the group law $\Lambda(\lambda_2)\Lambda(\lambda_1)=\Lambda(\lambda_2+\lambda_1)$. 
In view of the dependence of the group parameter on the spacetime coordinates $x^\mu$, this group is referred to as the local $U(1)$ 
gauge group. 

Local $U(1)$ transformations \eqref{2.2.1} are not a symmetry of the Lagrangian density \eqref{2.2}. A direct calculation gives
\begin{widetext}
\begin{eqnarray}
\L(\psi,\psi^*,d_\mu\psi,d_\mu\psi^*)&\not=&\L(\Lambda(\lambda)\psi,\Lambda(\lambda)\psi^*,d_\mu\Lambda(\lambda)\psi,d_\mu\Lambda(\lambda)\psi^*)
\nonumber\\
&=&\frac{ic}{2}\psi^*(d_0\psi-i(d_0\lambda)\psi)-\frac{ic}{2}\psi(d_0\psi^*+i(d_0\lambda)\psi^*)
-\frac{1}{2m}\left(\nabla\psi^*+i(\nabla\lambda)\psi^*\right)\cdot
\left(\nabla\psi-i(\nabla\lambda)\psi\right)\nonumber\\
\label{2.2.2}
\end{eqnarray}
\end{widetext}
If we desire a Lagrangian density that is invariant under local $U(1)$ gauge transformations, then the structure of \eqref{2.2.2} implies that we may couple 
the Schr\"odinger matter field to another field so that the extraneous terms coming from the derivatives of the gauge parameter $\lambda$  are canceled 
by suitable terms arising from the gauge transformations of this second field. The simplest choice is to couple $d_\mu\psi$ and $d_\mu\psi^*$ to a real field $A_\mu(x^\mu)=(A_0(x^\mu), \bs{A}(x^\mu))$, where, under rotations $A_0$ is a scalar and $\bs{A}$ is a vector,  with the requirement 
\begin{eqnarray}
(\Lambda(\lambda)A_0)(x^\mu)&=&A_0(x^\mu)+\frac{1}{g}d_0\lambda(x^\mu)\label{2.2.3}\\
(\Lambda(\lambda)\bs{A})(x^\mu)&=&\bs{A}(x^\mu)+\frac{1}{g}\nabla\lambda(x^\mu)\label{2.2.4}
\end{eqnarray}
where $g$ is a constant. From these equations and \eqref{2.2.2}, we see that terms $(d_0+igA_0)\psi$ and $(\nabla+ig\bs{A})\psi$, as well as their complex conjugates transform under local gauge transformations exactly as $\psi$ and $\psi^*$ in \eqref{2.2.1}. Therewith, as the Lagrangian density invariant 
under local gauge transformations 
\eqref{2.2.1}, \eqref{2.2.3} and \eqref{2.2.4}, we have
\begin{eqnarray}
\L&=&\frac{ic}{2}\psi^*(d_0+igA_0)\psi-\frac{ic}{2}\psi(d_0-igA_0)\psi^*\nonumber\\
&&\quad -\frac{1}{2m}\left(\nabla+ig\bs{A}\right)\psi\cdot\left(\nabla-ig\bs{A}\right)\psi^*\label{2.2.5}
\end{eqnarray}

The Euler-Lagrange equations obtained from \eqref{2.2.5} are
\begin{eqnarray}
ic(d_0+igA_0)\psi&=&-\frac{1}{2m}(\nabla+ig\bs{A})^2\psi\nonumber\\
-ic(d_0-igA_0)\psi^*&=&-\frac{1}{2m}(\nabla-ig\bs{A})^2\psi^*\label{2.2.6}
\end{eqnarray}
In the limit where the coupling constant $g$ vanishes, these equations reduce to the familiar free particle Schr\"odinger equation and its complex conjugate, while \eqref{2.2.5} reduces to the free particle Lagrangian density \eqref{2.1}.

It follows from the transformation formulas \eqref{2.2.3} and \eqref{2.2.4} that the linear combinations of derivatives $(\nabla A_0-d_0\bs{A})$ and $\nabla\times{\bs{A}}$ are also invariant under local $U(1)$ transformations. Therefore, we may add any scalar function 
of  $(\nabla A_0-d_0\bs{A})$ and $(\nabla\times{\bs{A}})$ to \eqref{2.2.5} so as to preserve local $U(1)$ invariance. To that end, define
\begin{eqnarray}
{\bs{E}}(x^\mu)&:=&(\nabla A_0-d_0{\bs{A}})(x^\mu)\label{2.2.10}\\
c{\bs{B}}(x^\mu)&:=&(\nabla\times{\bs{A}})(x^\mu)\label{2.2.11}
\end{eqnarray}
Then, the most general $U(1)$-invariant Lagrangian density  is 
\begin{eqnarray}
\L&=&\frac{ic}{2}\psi^*(d_0+igA_0)\psi-\frac{ic}{2}\psi(d_0-igA_0)\psi^*\nonumber\\
&&-\frac{1}{2m}\left(\nabla+ig\bs{A}\right)\psi\cdot\left(\nabla-ig\bs{A}\right)\psi^*       +f(\bs{E},\bs{B})\nonumber\\
\label{2.2.7}
\end{eqnarray}
where $f$ is an arbitrary sufficiently well behaved real-valued scalar (under rotations) function of  $\bs{E}=\nabla A_0-d_0{\bs{A}}$ and $\bs{B}=\frac{1}{c}\nabla\times\bs{A}$. 

The Euler-Lagrange equations for $\psi$ and $\psi^*$ obtained from the Lagrangian density \eqref{2.2.7} are the same as \eqref{2.2.6}.
The Euler-Lagrange equation for  $A_0$ is 
\begin{equation}
d_i\left(\frac{\partial f}{\partial E_i}\right)=-gc\psi^*\psi\label{2.2.8}
\end{equation}
and that for $A_i$ is
\begin{eqnarray}
-d_0\left(\frac{\partial f}{\partial E_i}\right)-\epsilon_{ijk}\frac{1}{c}d_j\left(\frac{\partial f}{\partial B_k}\right)&=&-\frac{ig}{2m}\left(\psi(d_i-igA_i)\right.\psi^*\nonumber\\
&&-\left.\psi^*(d_i+igA_i)\psi\right)\nonumber\\
\label{2.2.9}
\end{eqnarray}

For the sake of notational economy, we define 
\begin{eqnarray}
\nabla_{\bs{E}}f&:=&\left(\frac{\partial f}{\partial E_1}, \frac{\partial f}{\partial E_2},\frac{\partial f}{\partial E_3}\right)\nonumber\\
\nabla_{\bs{B}}f&:=&\left(\frac{\partial f}{\partial B_1}, \frac{\partial f}{\partial B_2},\frac{\partial f}{\partial B_3}\right)\label{2.2.9c}
\end{eqnarray}
In terms of these definitions, the equations of motion \eqref{2.2.8} and \eqref{2.2.9} can be written as vector equations:
\begin{eqnarray}
\nabla\cdot\nabla_{\bs{E}}f&=&-gc\psi^*\psi\label{2.2.8b}\\
-d_0\nabla_{\bs{E}}f-\frac{1}{c}\nabla\times\nabla_{\bs{B}}f&=&
-\frac{ig}{2m}\left(\psi(d_i-igA_i)\psi^*\right.\nonumber\\
&&-\left.\psi^*(d_i+igA_i)\psi\right)\nonumber\\
\label{2.2.9b}
\end{eqnarray}

\subsection{Maxwell's equations}

The definitions \eqref{2.2.10} and \eqref{2.2.11} of $\bs{E}$ and $\bs{B}$ trivially imply the homogeneous Maxwell equations:
\begin{eqnarray}
\nabla\cdot\bs{B}&=&0\nonumber\\
\nabla\times\bs{E}+d_0 c\bs{B}&=&\nabla\times\bs{E}+\frac{d}{dt}\bs{B}=0\label{2.2.6c}
\end{eqnarray}

The two inhomogeneous Maxwell equations can be obtained as a special case of \eqref{2.2.8b} and \eqref{2.2.9b}. In particular,  let 
\begin{equation}
f(\bs{E},\bs{B})=\frac{c}{2}\left(\bs{E}^2-\bs{B}^2\right)\label{2.2.6d}
\end{equation}
and define charge and current densities by
\begin{eqnarray}
\rho(x^\mu)&:=&-g\psi^*(x^\mu)\psi(x^\mu)\label{2.2.12}\\
\bs{j}(x^\mu)&:=&-\frac{ig}{2m}\left\{\psi(x^\mu)(\nabla-ig\bs{A})\psi^*(x^\mu)\right.\nonumber\\
&&\ -\left.\psi^*(x^\mu)(\nabla+ig\bs{A})\psi(x^\mu)\right\}
\label{2.2.13}
\end{eqnarray}
Then, \eqref{2.2.8b} and \eqref{2.2.9b} reduce to the inhomogeneous Maxwell equations:
\begin{eqnarray}
(\nabla\cdot{\bs{E}})(x^\mu)&=&\rho(x^\mu)\label{2.2.14}\\
(\nabla\times{\bs{B}})(x^\mu)-\left(\frac{d}{dt}{\bs{E}}\right)(x^\mu)&=&{\bs{j}}(x^\mu)\label{2.2.15}
\end{eqnarray}
It follows from the matter field equations of motion \eqref{2.2.6} that the continuity equation holds for $\rho$ and $\bs{j}$:
\begin{equation}
\frac{d}{dt}\rho+\nabla\cdot{\bs{j}}=0\label{2.2.16}
\end{equation}
Therefore, it is consistent to interpret $\rho$ and $\bs{j}$ as the charge and current densities, respectively.  
\vskip 0.5cm

 An analogous calculation can be carried out for the classical Klein-Gordon field \cite{ryder}. Here, too, we find that invariance under local $U(1)$ gauge transformations requires that the matter field be coupled to a gauge field $A_\mu$ with the transformation properties \eqref{1.3}, the covariant form of  \eqref{2.2.3} and \eqref{2.2.4}. Again, Maxwell's equations arise  as the equations of motion for $A_\mu$ when $f$ is of the form \eqref{2.2.6d}. 

Maxwell's equations are relativistic equations. Therefore, as far as spacetime transformation properties are concerned, it appears mathematically tenable and physically sensible that Maxwell's equations are coupled to the relativistically invariant Klein-Gordon equation \cite{ryder}. Then, this relativistic result and the preceding non-relativistic result show that the invariance under local $U(1)$ gauge transformations requires that 
a matter field must couple to the electromagnetic field, \emph{whether the matter field is relativistic or  non-relativistic}. Our result is reminiscent of the Feynman-Dyson result \cite{dyson}, albeit it is derived from an entirely different approach.  In a certain sense, these results imply that the structure of Maxwell's equations is so not because of structure of spacetime but because of the invariance under $U(1)$ transformations. 

Both our result and the Feynman-Dyson result raise the question whether the Galilean invariance of the free particle equations motion infer the Galilean invariance of the gauge field equations. If this is not true, it is illuminating to consider at what point of the derivation does this invariance is broken. To further examine these  questions, we must consider the transformation of the equations \eqref{2.2.14}-\eqref{2.2.16} as well as \eqref{2.2.6} under Galilean transformations, our task in the next Section. 

\section{Galilean transformations of gauge field equations}\label{sec3}

We define the Galilei group $\G$, the symmetry group of non-relativistic spacetime, by the transformation rules
\begin{eqnarray}
\tilde{t}&=&t+\tt{b}\nonumber\\
\tilde{\bf{x}}&=&R{\bf{x}}+{\bf{v}}t+{\bf{a}}\label{3.1}
\end{eqnarray}
where $R$ is an orthogonal rotation matrix, ${\bf{a}}$ is a space translation, $\tt{b}$ is a time translation, 
and $\bs{v}$ is a pure Galilei transformation. In the notation introduced in Section \ref{sec2} for the spacetime vectors $x^\mu=(ct,\bs{x})$, 
the transformation rules \eqref{3.1} can be rewritten as 
\begin{eqnarray}
\tilde{x}^0&=&x^0+b\nonumber\\
\tilde{\bs{x}}&=&R\bs{x}+\bs{\beta}x^0+\bs{a}\label{3.1a}
\end{eqnarray}
where $b=c\tt{b}$, $\bs{\beta}=\frac{\bs{v}}{c}$. 
If we parametrize  group elements by $g(b,\bs{a},\bs{\beta},R)$, then from \eqref{3.1},
\begin{eqnarray}
&&g(b_2,{\bs{a}_2},{\bs{\beta}_2},R_2)g (b_1,{\bf{a}_1},{\bs{\beta}_1},R_1)\nonumber\\
&&\ =g(b_1+b_2, \bs{a}_2+R_2{\bf{a}_1}+b_1{\bs{\beta}_2},{\bs{\beta}_2}+R_2{\bs{\beta}_1}, R_2 R_1)\nonumber\\
&&g^{-1}(b,{\bs{a}},{\bs{\beta}},R)=g(-b, -R^{-1}({\bs{a}}-b{\bs{\beta}}), -R^{-1}{\bs{\beta}}, R^{-1})\nonumber\\
\label{3.2}
\end{eqnarray}
The product rule \eqref{3.2} shows  each  
 $g\in\G$ has the  unique decomposition
\begin{equation}
g(b,\bs{a},\bs{\beta},R)=g(b,\bs{a},0,I)g(0,0,\bs{\beta},R)\label{3.3}
\end{equation}
in terms of the elements of the subgroup of translations and that of homogeneous Galilean transformations. 
We will denote the group of homogeneous Galilei transformations by $\tilde{\G}$ and its elements by $\tilde{g}$. 

The group $\tilde{\G}$  has a natural representation 
by $4\times4$-matrices:
\begin{equation}
D(\tilde{g})=\left(
\begin{array}{cc}
1&0\\
\bs{\beta}&R
\end{array}
\right)\label{3.4}
\end{equation}
Recall that if $D$ is a matrix representation of any group $G$ then we may define its dual representation $C$ by
\begin{equation}
C(g)=D^\dagger(g^{-1})\label{3.5}
\end{equation}
It is straightforward to verify the product law $C(g_2)C(g_1)=C(g_2g_1)$. 
For the homogeneous Galilei group with $D$ given by \eqref{3.4}, we have 
\begin{equation}
C(\tilde{g})=\left(
\begin{array}{cc}
1&-\widehat{R^{-1}\bs{\beta}}\\
0&R
\end{array}
\right)\label{3.6}
\end{equation}
where $\widehat{R^{-1}\bs{\beta}}$ is the dual (row) vector that corresponds to $R^{-1}\bs{\beta}$ in ${\mathbb{R}}^3$. 

The dual space of $\mathbb{R}^4$ (with respect to the usual inner product $x\cdot y=x_0y_0+x_1y_1+x_2y_2+x_3y_3$) is isomorphic to 
$\mathbb{R}^4$ and can be identified with  itself. However, since the elements of the dual space transform under the representation 
\eqref{3.6}, we denote its elements with the subscript $x_\mu$. That is,
\begin{equation}
\tilde{x}_\mu=\left(C(\tilde{g})x\right)_\mu=\sum_{\nu=0}^3C(\tilde{g})_{\mu\nu}x_\nu\label{3.7}
\end{equation}
whereas 
\begin{equation}
\tilde{x}^{\mu}=\left(D(\tilde{g})x\right)^\mu=\sum_{\nu=0}^3D(\tilde{g})_{\mu\nu}x^\nu\label{3.8}
\end{equation}

Admittedly, our notation is analogous to that of special and general relativity where superscript and subscript indices are used to 
distinguish between vectors transforming contravariantly and covariantly.  We emphasize that here 
we are working in a flat four dimensional Euclidean space, and the only difference between the 
vectors $x^\mu$ and $x_\mu$ is that they furnish the representations $D$ and $C$ of the 
homogeneous Galilei group, respectively. 

In component form, the transformation rule of a vector $\tilde{x}_\mu$ under $C$ is
\begin{eqnarray}
\tilde{x}_0&=&x_0-R^{-1}\bs{\beta}\cdot\bs{x}\nonumber\\
\tilde{\bs{x}}&=&R\bs{x}\label{3.9}
\end{eqnarray}

Equations \eqref{3.9} have no obvious physical interpretation as a transformation law for spacetime.  However, as noted by Le Bellac and L\'evy-Leblond \cite{bellac}, transformation formulas \eqref{3.1a} and \eqref{3.9} are the two fundamental small $\bs{\beta}$ limits of Lorentz transformations. In fact,  Lorentz transformation formulas approximated to first order in $\bs{\beta}$ (so that $\gamma\approx1$) do not define a group at all. For instance, suppressing rotations and taking the boost parameter $\bs{\beta}$ to be entirely in the $x_1$-direction we have, to first order in $\beta$,  
\begin{eqnarray}
\tilde{x}_0&=&x_0-\beta x_1\nonumber\\
\tilde{x}_1&=&x_1-\beta x_0\label{3.10}
\end{eqnarray}
These are clearly not the standard Galilean transformation rules, and \eqref{3.10} does not define a group. 
The two sets of transformation rules \eqref{3.1a} and \eqref{3.9}, i.e.~the representations $D$ and $C$ of 
$\tilde{\cal G}$, follow from the limiting case  \eqref{3.10} \emph{and} the supplementary condition $|x_0|\geq|\bs{x}|$ or 
$|x_0|\leq|\bs{x}|$, respectively.  

It follows that the Euclidean inner product $x\cdot y=\sum_\mu x^\mu y_\mu=x^0y_0+x^1y_1+x^2y_2+x^3y_3$ is invariant under homogeneous Galilean transformations. Furthermore, if $x^\mu$ transform under $D$ then the differential operators $\frac{d}{dx^\mu}\equiv d_\mu$ transform under $C$:
\begin{eqnarray}
\tilde{d}_0\equiv\frac{d}{d{\tilde{x}}^0}&=&d_0-R^{-1}\bs{\beta}\cdot\nabla\nonumber\\
\tilde\nabla&=&R\nabla\label{3.11}
\end{eqnarray}
The transformation properties \eqref{3.1a} and \eqref{3.11} were the reason 
that in Section 2 we denoted the spacetime vectors with superscript indices and differential operators with subscript indices. 

\subsection{The Schr\"odinger equation and unitary representations of the Galilei group}\label{sec3.1}

Following Wigner's pioneering study of the Poincar\'e group \cite{wigner}, it has been known that quantum mechanically relevant representations of a symmetry group are  unitary and, in general, projective. Recall that a unitary projective representation of a Lie group $G$ in a Hilbert space $\cal H$ is a mapping ${\cal H}\otimes G\to{\cal H}$ such that the unitary operators $U(g)$, $g\in G$,  fulfill the identity
\begin{equation}
U(g_2)U(g_1)=e^{-i\omega(g_2,g_1)}U(g_2g_1)\label{3.1.1}
\end{equation}
where $\omega(g_2,g_1)$ is a real valued function determined by the structure of the group. For a true representation, $e^{i\omega(g_2,g_1)}=1$. As shown by Bargmann \cite{bargmann}, for many relevant symmetry groups, such as the rotation group, Lorentz group and Poincar\'e group, unitary projective representations are equivalent to true representations of their universal covering groups.  In contrast, there are infinitely many classes of projective representations of the Galilei group that are not equivalent to true representations of its covering group, and it is precisely these representations that are physically relevant \cite{bargmann}. For instance, it is only for these projective representations that position operators can be defined (as generators of Galilean boosts) \cite{hammermesh}. For a review of the Galilei group and its physically meaningful representations, see \cite{levy-leblond}. 

The scalar function $\omega(g_2,g_1)$ on the group manifold defining the projective representations \eqref{3.1.1} can be determined from the product rule of the  group. For the  Galilei group,
\begin{equation}
\omega(g_2,g_1)=\frac{1}{2}mc\left(\bs{a}_2\cdot{R_2\bs{\beta}_1}-\bs{\beta}_2\cdot{R_1\bs{a}_1}+b_1\bs{\beta}_2\cdot{R_2\bs{\beta}_1}\right)\label{3.1.2}
\end{equation}
where $m$ is an arbitrary real number \cite{bargmann,levy-leblond}. 

Wigner's study of the Poincar\'e group \cite{wigner} showed that an elementary relativistic quantum physical system, characterized by mass and spin, is associated to a unique 
unitary irreducible representation (UIR) of the Poincar\'e group. Mass and spin values of the physical system arise as the eigenvalues of the two Casimir operators of the representation. One might then expect that an elementary non-relativistic quantum physical system is similarly associated to a unitary irreducible representation of the the Galilei group. This is indeed the case, with the qualification that unitary irreducible representations be projective. 
More precisely, the Hilbert space of states of an elementary non-relativistic quantum system furnishes a unitary, irreducible, 
projective representation of the Galilei group. 

If $U$ is a unitary representation of a Lie group in a Hilbert space $\cal H$,  
then the differential of $U$ evaluated at the group identity, $\left.idU\right|_{e}$, furnishes a 
representation of the Lie algebra of the group by self-adjoint, generally unbounded operators.  If $U$ is such a representation of the 
Galilei group then a basis for the operator Lie algebra $\left.idU\right|_{e}$ can be chosen to consist of the generators of spacetime translations $\bs{P}$ and $H$, 
the generators of rotations $\bs{J}$, and the generators of Galilean velocity boosts $\bs{K}$. The mass operator associated with projective unitary representations can be adjoined to the enveloping algebra  of this operator Lie algebra as a central element. The operators $H$, $\bs{P}$, $\bs{J}$, $\bs{K}$, and $M$ fulfill the commutation relations
\begin{eqnarray}
\left[H,P_i\right]=\left[H,J_i\right]=0&&\left[H,K_i\right]=iP_i\nonumber\\
\left[P_i,P_j\right]=\left[K_i,K_j\right]=0&&\left[J_i,J_j\right]=i\epsilon_{ijk}J_k\nonumber\\
\left[J_i,P_j\right]=i\epsilon_{ijk}P_k &&\left[J_i,K_j\right]=i\epsilon_{ijk}K_k\nonumber\\
\left[K_i,P_j\right]&=&i\delta_{ij}M\label{3.1.3}
\end{eqnarray}
and $M$ commutes with all the operators. 

The Hilbert space $\cal H$ in which the Galilean Lie algebra \eqref{3.1.3} is defined as self-adjoint operators can be realized as the space of $L^2$-functions on the Cartesian product of the spectra of a complete system of commuting operators (CSCO) chosen from the associative enveloping algebra of \eqref{3.1.3}. All CSCO are equally good, and it is natural to choose one to include all the Casimir operators of the algebra. The commutation relations \eqref{3.1.3} show that the operators $W=H-\frac{1}{2M}{\bs{P}^2}$, $\bs{S}^2=\left(\bs{J}-\frac{1}{M}\bs{K}\times\bs{P}\right)^2$, and, trivially, $M$ all commute with all the Galilean generators. These invariant operators have physical interpretation as  internal energy, the square of total spin, and mass, respectively. In an irreducible unitary representation, each is proportional to the identity:  $W=wI$, $M=mI$, and $\bs{S}^2=s(s+1)I$ with constant values for $w,\ m\in{\mathbb{R}}$ and $s=0,1/2,1,\cdots$.  This means that the set of values $(w,s,m)$ uniquely determine projective UIR's of the Galilei group. Therefore, we denote the representation Hilbert space by ${\cal H}(w,s,m)$. When dealing with a single UIR, it is possible to set $w=0$ without loss of generality. 

A common choice for a CSCO is $\bs{P}$, $S_3$, $W$, $\bs{S}^2$, and $M$.  Here, $S_3$ is the third component of a suitable spin vector operator $\bs{S}$. It should be noted that while the operator $\bs{S}^2$ is uniquely defined in a UIR, there are infinitely many vector operators $\bs{S}$ satisfying the characteristic commutation relations $\left[S_i,S_j\right]=i\epsilon_{ijk}S_k$ such that $\bs{S}^2=s(s+1)I$. The point spectrum of any such $S_3$ is 
$-s,-s+1,\cdots,s-1,s$. 

The representation Hilbert space ${\cal H}(w,s,m)$ 
can now be realized as the space of functions $L^2({\mathbb{R}}^3)\otimes{\mathbb{C}}^{(2s+1)}$. We denote the values of 
these functions by $\psi(\bs{p},s_3)$ where $\bs{p}\in\mathbb{R}^3$ and 
$s_3=-s,-s+1,\cdots,s-1,s$. The method of induced representations developed by Wigner and Mackey can be used to 
construct the unitary operators $U(g)$ for the representation $\G\otimes{\cal H}(w,s,m)
\to{\cal H}(w,s,m)$, see, for instance, \cite{levy-leblond}. Here, we quote the result: 
\begin{eqnarray}
\left(U(g)\psi\right)(\bs{p},s_3)&=&\exp{\left[-i\left(\frac{1}{2}mc\bs{a}\cdot\bs{\beta}+\bs{a}\cdot\bs{p}'-bE'\right)\right]}\nonumber\\
&&\times\sum_{\s_3'}D^{s}({\mathcal{R}}(({\bf p}, E),\tilde{g}))_{s_3s_3'}\psi(\bs{p}',s_3')\nonumber\\
\label{3.1.4}
\end{eqnarray}
where  $E=\frac{\bs{p}^2}{2m}+w$ and 
$(\bs{p}',E')=\tilde{g}^{-1}\left(\bs{p},E\right)=(R^{-1}(\bs{p} - mc{\bs \beta}),
 E + c{\bs\beta}\cdot\bs{p} + 
\frac{1}{2}mc^2{\bs\beta}^2)$. As defined by \eqref{3.3}, $\tilde{g}=g(0,0,{\bs{\beta}},R)$ is the homogeneous Galilean transformation associated with 
$g(b,{\bf{a}},{\bs{\beta}},R)$.  The $D^s$ are the $2s+1$-dimensional unitary matrices and their arguments 
${\mathcal{R}}(({\bf p}, E),{\tilde{g}})$ are elements of the ``little group" of $\mathcal{G}$ for a massive particle. 
Recall that the little group is the largest subgroup that leaves a standard momentum-energy pair  $({\bf{p}}_0,E_0)$ invariant. 
For a massive particle, the little group of both the Galilean and Poincar\'e groups is isomorphic to the rotation group, and therefore, the 
$D^s({\mathcal{R}}(({\bf{p}},E),\tilde{g}))$ is simply the unitary $2s+1$ dimensional representation of the rotation group. 
By definition, the little group depends on the choice of $({\bf{p}}_0,E_0)$ 
which is arbitrary aside from the constraint $E-\frac{1}{2m}{\bf{p}}^2=w$. 
However, all of these different choices lead to equivalent representations of $\mathcal{G}$~\cite{levy-leblond}, 
and therefore we may use any momentum-energy pair $({\bf{p}},E)$ to construct the general expression for the 
representation. The choice $({\bf{0}},w)$ is particularly simple in that ${\mathcal{R}}(({\bf 0}, w),{\tilde{g}})=R$, i.e., 
the little group of $\mathcal{G}$ can be chosen to be $SU(2)$, independently of the momentum and energy of the particle. 
Then, \eqref{3.5} becomes
\begin{eqnarray}
(U(g)\psi)(\bs{p},s_3) &=& e^{-i\left(\frac{1}{2}mc \mathbf{a}\cdot\bs{\beta} +\bs{a}\cdot\bs{p}' -bE'\right)} \times\nonumber\\
&&\sum_{s'_3}D^{s}(R)_{s_3s_3'}\psi(\bs{p}',s_3'),
\label{3.1.5}
\end{eqnarray} 
and for a spinless particle of mass $m$, we simply have 
\begin{equation}
(U(g)\psi)(\bs{p}) = e^{-i\left(\frac{1}{2}mc \mathbf{a}\cdot\bs{\beta} +\bs{a}\cdot\bs{p}' -bE'\right)} \psi(\bs{p}').
\label{3.1.6}
\end{equation} 
Along with \eqref{3.1.2},  it is straightforward to verify that each of \eqref{3.1.4}, \eqref{3.1.5} and \eqref{3.1.6} 
defines a projective representation of the Galilei group. In order to avoid inessential complications, we will henceforth consider only the spinless case \eqref{3.1.6} and its position representation,  defined below in \eqref{3.1.7}. 

From the last equality of  the commutation relations \eqref{3.1.3}, we see that the vector operator $\bs{Q}=\frac{1}{M}\bs{K}$ satisfies the canonical Heisenberg commutation relations with the momentum operator $\bs{P}$. Therefore, a position operator $\bs{Q}$ exists in the Galilean algebra extended by the central charge $M$. We may choose the set $\{\bs{Q},S_3, W, S^2, M\}$ as a CSCO to obtain the position wavefunctions $\psi\in L^2({\mathbb{R}}^3)\otimes{\mathbb{C}}^{(2s+1)}$. For a spinless particle, $\psi\in L^2({\mathbb{R}}^3)$.  

If $U$ is a unitary irreducible representation of the Galilean group, it is common to denote the time translated functions $(U(t,0,0,I)\psi)(\bs{x})=\langle\bs{x}|U(t,0,0,I)\psi\rangle$ by 
$\psi(\bs{x},t)$ or in the notation of Section \ref{sec2}, by $\psi(x^\mu)$ or more simply by $\psi(x)$. The unitary operators $U(g)$ furnishing the spinless 
projective UIR of the Galilei group are defined by following transformation rule for $\psi(x)$:
\begin{equation}
\psi'(x)\equiv(U(b,{\bs{a}},{\bs{\beta}},R)\psi)(x)=e^{-i\gamma(x')}\psi(x')\label{3.1.7}
\end{equation}
where 
\begin{equation}
\gamma(x')=mc\left(-R^{-1}{\bs{\beta}}\cdot{\bs{x}'}+\frac{1}{2}{\bs{\beta}^2}x'^0-C\right)\label{3.1.7b}
\end{equation}
The integration constant $C=-\frac{1}{2}\bs{a}\cdot\bs{\beta}+\frac{1}{2}b\bs{\beta}^2$ and  $x'$ is defined by the inverse transformation of \eqref{3.1a}, i.e., 
\begin{eqnarray}
x'^{0}&=&x^0-b\nonumber\\
{\bs{x}}'&=&R^{-1}\bs{x}-(R^{-1}\bs\beta)x^0-R^{-1}(\bs{a}-b\bs{\beta})\label{3.1.8}
\end{eqnarray}
Along with \eqref{3.1.2}, \eqref{3.1.7b}, and \eqref{3.1.8}, it is straightforward to verify that \eqref{3.1.7} defines a projective UIR of the Galilei group.  
It is necessary to define $x'$ as the inverse group element in order to ensure that \eqref{3.1.7} defines a homomorphism $\G\to U(\G)$. 

It is significant that the $L^2({\mathbb{R}}^3)$-functions transforming under $\G$ as in \eqref{3.1.7} are also the solutions to the Schr\"odinger equation 
for a free particle of mass $m$ \cite{levy-leblond}. 
In fact, the transformation formula \eqref{3.1.7} can be derived by demanding that the Schr\"odinger equation, or 
or equivalently,  Lagrangian density \eqref{2.2} be invariant in form under Galilean transformations.   In order to show its Galilean invariance, we first write \eqref{2.2} in terms of the transformed fields \eqref{3.1.7}:
\begin{equation}
\L'_{\rm free}=\frac{ic}{2}\psi'^* d_0\psi'-\frac{ic}{2}\psi' d_0 \psi'^*-\frac{1}{2m}\nabla\psi'\cdot\nabla\psi'^*\label{3.1.9}
\end{equation}
Since the arguments of the functions on the right hand side of \eqref{3.1.7} are $x'^\mu$, we must express the differential operators $d_\mu$ in \eqref{3.1.9} in terms 
of the transformed operators $d'_\mu$. It follows from \eqref{3.1.8},
\begin{equation}
d_\mu=C(\tilde{g})d'_\mu\label{3.1.10}
\end{equation}
or, in the component form
\begin{eqnarray}
d_0&=&d'_0-R^{-1}{\bs{\beta}}\cdot\nabla'\nonumber\\
\nabla&=&R\nabla'\label{3.1.11}
\end{eqnarray}
Substituting  \eqref{3.1.7} and \eqref{3.1.10} into \eqref{3.1.9} and after a bit of algebra, we obtain
\begin{eqnarray}
\L'_{\rm free}&=&\frac{ic}{2}\psi^*(x') d'_0\psi(x')-\frac{ic}{2}\psi(x') d'_0 \psi^*(x')\nonumber\\
&&-\frac{1}{2m}\nabla'\psi(x')\cdot\nabla'\psi^*(x')\label{3.1.12}
\end{eqnarray}
Dropping the primes in spacetime coordinates and differential operators, we therefore have
\begin{equation}
\L'_{\rm free}=\L_{\rm free}\label{3.1.13}
\end{equation}
which ensures the invariance of the free particle Schr\"odinger equation under Galilean transformations. Thus, there is a one-to-one correspondence between the 
space of square integrable  solutions of the free particle Schr\"odinger equation and the projective UIR's of the Galilei group. In this sense, we may regard a projective UIR of $\G$ as the mathematical image of 
a free non-relativistic particle in much the same way a UIR of the Poincar\'e group is the mathematical image of a relativistic particle.

\subsection{Galilean transformations of the gauge invariant Lagrangian}\label{sec3.2}

We now consider the transformation properties of the full Lagrangian density \eqref{2.2.7} for the coupled matter and gauge fields under $\G$.  We already have the transformation rules for the matter field \eqref{3.1.7} and the differential operators \eqref{3.1.10}. It only remains to deduce  the transformation rules for the gauge fields $A_0$ and $\bs{A}$. The simplest non-trivial assumption is that $A_0$ and $\bs{A}$ are components of a Galilean vector field, i.e., a vector field under $\tilde{\cal G}$. From \eqref{3.7} and \eqref{3.8}, we see that there are two different kinds of Galilean vector field. Therewith, we have two inequivalent transformation rules  for the gauge field: 
\begin{enumerate} 
\item The gauge field is a vector field under \eqref{3.6}, i.e., $A_\mu=(A_0,{A}_i)$ with
\begin{equation}
A'_\mu(x)=C(\tilde{g})_{\mu\nu}A_\nu(x')\label{3.2.1}
\end{equation}
In component form, this reads as
\begin{eqnarray}
A'_0(x)&=&A_0(x')-R^{-1}\bs{\beta}\cdot\bs{A}(x')\nonumber\\
\bs{A}'(x)&=&R\bs{A}(x')\label{3.2.4}
\end{eqnarray}
With \eqref{3.2.4} for $A_\mu$, the electric and magnetic fields transform as 
\begin{eqnarray}
\bs{E}'(x)&=&\nabla A'_0(x)-d_0\bs{A}'(x)\nonumber\\
&=&R\bs{E}(x')-\bs{\beta}\times cR\bs{B}(x')\nonumber\\
\bs{B}'(x)&=&\nabla\times\bs{A}'(x)=R\bs{B}(x')\label{3.2.7}
\end{eqnarray}

\item The gauge field is a vector field under \eqref{3.4}, i.e., $A^\mu=(A^0,{A}^i)$ with 
\begin{equation}
A'^\mu(x)=D(\tilde{g})_{\mu\nu}A^\nu(x')\label{3.2.2}
\end{equation}
In component form, this reads as 
\begin{eqnarray}
A'^0&=&A^0(x')\nonumber\\
\bs{A}'(x)&=&R\bs{A}(x')+\bs{\beta}A_0(x')\label{3.2.2a}
\end{eqnarray}
With \eqref{3.2.2a} for $A^\mu$, the electric and magnetic fields transform as
\begin{eqnarray}
\bs{E}'(x)&=&\nabla A'^0(x)-d_0\bs{A}'(x)=R\bs{E}(x')\nonumber\\
c\bs{B}'(x)&=&\nabla\times{\bs{A}}'(x)=cR\bs{B}(x')+\bs{\beta}\times R\bs{E}(x')\nonumber\\
\label{3.2.2b}
\end{eqnarray}
\end{enumerate}
In  \eqref{3.2.1}-\eqref{3.2.2b}, the $x'$ is defined by the inverse Galilean transformation \eqref{3.1.8}. 

We see from \eqref{3.1.10} that when $x^\mu$ transforms as in \eqref{3.1.8}, $x'^\mu=(g^{-1}x)^\mu$, the differential operators transform as a vector under the dual representation $C(\tilde{g})$. Since the gauge field $(A_0,\bs{A})$ is coupled to the  differential operators $(d_0,\nabla)$ in the matter part of the Lagrangian density \eqref{2.2.7}, the transformation rule \eqref{3.1.10} for $d_\mu$ suggests that the gauge field must transform as in \eqref{3.2.1} rather than \eqref{3.2.2}.  
With this choice for $A_\mu$, the electric and magnetic fields transform as \eqref{3.2.7}. 

Equations \eqref{3.2.7} can be obtained as the non-relativistic limit of the relativistic transformation rules for the electromagnetic tensor, subject to the constraint $c|\bs{B}|>>|\bs{E}|$ \cite{bellac}. Hence, \eqref{3.2.7} are often called the `magnetic limit' of the relativistic transformation formula. Similarly, \eqref{3.2.2b}, which follows as the non-relativistic limit when $c|\bs{B}|<<|\bs{E}|$ is called the `electric limit'. However, it should be remarked  that our  transformations rules for the gauge field or the electromagnetic field do not entail any limiting case of a relativistic vector or tensor field. We are working purely in a non-relativistic setting with tools from representation and group theory.  

After these preliminary observations, we now consider the Lagrangian density \eqref{2.2.7} in terms of the transformed matter and gauge fields:
\begin{eqnarray}
\L'&=&\frac{ic}{2}\psi'^*(d_0+igA'_0)\psi'-\frac{ic}{2}\psi'(d_0-igA'_0)\psi'^*\nonumber\\
&&-\frac{1}{2m}\left(\nabla+ig\bs{A}'\right)\psi'\cdot\left(\nabla-ig\bs{A}'\right)\psi'^*\nonumber\\
&&\ +f\left((\nabla A'_0-d_0\bs{A}'), (\nabla\times{\bs{A}}')\right)\label{3.2.8}
\end{eqnarray}
It is convenient to split the full Lagrangian density  into two parts  $\L=\L_{\rm m}+\L_{\rm g}$ such that 
\begin{widetext}
\begin{eqnarray}
\L'_{\rm m}&=&\frac{ic}{2}\psi'^*(d_0+igA'_0)\psi'-\frac{ic}{2}\psi'(d_0-igA'_0)\psi'^*
-\frac{1}{2m}\left(\nabla+ig\bs{A}'\right)\psi'\cdot\left(\nabla-ig\bs{A}'\right)\psi'^*\label{3.2.9}\\
\L'_{\rm g}&=&f\left((\nabla A'_0-d_0\bs{A}'), (\nabla\times{\bs{A}}')\right)=f(\bs{E}',\bs{B}')\label{3.2.10}
\end{eqnarray}
\end{widetext}

The matter field Lagrangian density is invariant in form under Galilean transformations. To see this, let us substitute \eqref{3.1.7} and its complex conjugate for $\psi'$ and $\psi'^*$. Likewise,  \eqref{3.2.1} for $A'_\mu$. Then,
\begin{widetext}
\begin{eqnarray}
\L'_{\rm m}&=&\frac{ic}{2}e^{i\gamma(x')}\psi^*(x')\left(d_0+ig(A_0(x')-R^{-1}\bs{\beta}\cdot\bs{A}(x'))\right)e^{-i\gamma(x')}\psi(x')\nonumber\\
&&-\frac{ic}{2}e^{-i\gamma(x')}\psi(x')\left(d_0-ig(A_0(x')-R^{-1}\bs{\beta}\cdot\bs{A}(x'))\right)e^{i\gamma(x')}\psi^*(x')\nonumber\\
&&-\frac{1}{2m}\left(\nabla+igR\bs{A}(x')\right)e^{-i\gamma(x')}\psi(x')
\cdot\left(\nabla-igR\bs{A}(x')\right)e^{i\gamma(x')}\psi^*(x')\label{3.2.11}
\end{eqnarray}
\end{widetext}
Since the arguments of the functions are $x'$, in order to carry out the differentiations, we must substitute for $d_\mu$ from \eqref{3.1.10}. Then, after a bit of algebra, we get 
\begin{widetext}
\begin{eqnarray}
\L'_{\rm m}&=&\frac{ic}{2}\psi^*(x')(d'_0+igA_0(x'))\psi(x')-\frac{ic}{2}\psi(x')(d_0-igA_0(x'))\psi^*(x')
-\frac{1}{2m}\left(\nabla'+ig\bs{A}(x')\right)\psi(x')\cdot\left(\nabla'-ig\bs{A}(x')\right)\psi^*(x')\nonumber\\
&&+\left(cd'_0\gamma-c\bs{\beta}\cdot R\nabla'\gamma-\frac{1}{2m}\nabla'\gamma\cdot\nabla'\gamma\right)\psi^*(x')\psi(x')
+i\Bigl(\frac{1}{2m}R\nabla'\gamma+\frac{1}{2}c\bs{\beta}\Bigr)\cdot\Bigl(\psi R(\nabla'-ig\bs{A})\psi^*-\psi^*(\nabla'+ig\bs{A})\psi\Bigr)\nonumber\\
\label{3.2.12}
\end{eqnarray}
\end{widetext}
From \eqref{3.1.7b}, $\left(cd'_0\gamma-c\bs{\beta}\cdot R\nabla'\gamma-\frac{1}{2m}\nabla'\gamma\cdot\nabla'\gamma\right)=0$ and $\left(\frac{1}{2m}R\nabla'\gamma+\frac{1}{2}c\bs{\beta}\right)=0$. Therefore, the last two terms of \eqref{3.2.12} vanish, and we have
\begin{eqnarray}
\L'_{\rm m}&=&\frac{ic}{2}\psi^*(x')(d'_0+igA_0(x'))\psi(x')\nonumber\\
&&-\frac{ic}{2}\psi(x')(d_0-igA_0(x'))\psi^*(x')\nonumber\\
&&-\frac{1}{2m}\left(\nabla'+ig\bs{A}(x')\right)\psi(x')\cdot\left(\nabla'-ig\bs{A}(x')\right)\psi^*(x')\nonumber\\
\label{3.2.13}
\end{eqnarray}
A comparison of \eqref{3.2.13} with \eqref{2.2.7} shows that the matter field Lagrangian density $\L_{\rm m}$  remains invariant under Galilean transformations. Therewith, the equations of motion \eqref{2.2.6} are also Galilean invariant. 

An analogous calculation shows that the gauge field Lagrangian density $\L_{\rm g}$ is \emph{not} invariant under Galilean transformations.  As  in \eqref{3.2.11}, we 
first substitute for $A'_\mu$ with \eqref{3.2.1},
\begin{widetext}
\begin{equation}
\L'_{\rm g}=f\left((\nabla A'_0-d_0\bs{A}'),\nabla\times{\bs{A}'}\right)
=f\left\{[\nabla(A_0(x')-R^{-1}\bs{\beta}\cdot\bs{A}(x'))-d_0R\bs{A}(x')],\nabla\times R{\bs{A}}(x')\right\}
\end{equation}
\end{widetext}
and use \eqref{3.1.10} to carry out the differentiations. 
Or, equivalently, we may substitute for $\bs{E}'$ and $\bs{B}'$ with \eqref{3.2.7} in the second equality of \eqref{3.2.10}:
\begin{eqnarray}
\L'_{\rm g}&=&f(\bs{E}'(x),\bs{B}'(x))\nonumber\\
&=&f\{[R\bs{E}(x')-\bs{\beta}\times cR\bs{B}(x')],R\bs{B}(x')\}\nonumber\\
\label{3.2.14}
\end{eqnarray}
The expression \eqref{3.2.14} readily 
shows that $\L'_{\rm g}$ is not form invariant under Galilean transformations. Nevertheless,  possibly depending on the form of $f$ as a function of the field variables $\bs{E}$ and $\bs{B}$, it is possible for the equations of motion resulting from \eqref{3.2.14} to be the same as \eqref{2.2.8b} and \eqref{2.2.9b}. Therefore, we directly obtain the equations of motion from the full Lagrangian density for the transformed fields,
\begin{widetext}
\begin{eqnarray}
\L'&=&\frac{ic}{2}\psi^*(x')(d'_0+igA_0(x'))\psi(x')-\frac{ic}{2}\psi(x')(d_0-igA_0(x'))\psi^*(x')
-\frac{1}{2m}\left(\nabla'+ig\bs{A}(x')\right)\psi(x')\cdot\left(\nabla'-ig\bs{A}(x')\right)\psi^*(x')\nonumber\\
&&+f\left\{[\nabla' A_0(x')-d'_0\bs{A}(x')-R^{-1}\bs{\beta}\times \left(\nabla'\times\bs{A}(x')\right)],\left(\nabla'\times\bs{A}(x')\right)\right\}
\label{3.2.15}
\end{eqnarray}
\end{widetext}
For notational convenience, we define 
\begin{eqnarray}
\bs{\mathbb{E}}(x')&\equiv&\left(\nabla' A_0(x')-d'_0\bs{A}(x')-R^{-1}\bs{\beta}\times \left(\nabla'\times\bs{A}(x')\right)\right)\nonumber\\
&=&R^{-1}\bs{E}'(x)\nonumber\\
c\bs{\mathbb{B}}(x')&\equiv&\left(\nabla'\times \bs{A}(x')\right)=R^{-1}c\bs{B}'(x)
\end{eqnarray}
and write \eqref{3.2.15} as
\begin{widetext}
\begin{eqnarray}
\L'&=&\frac{ic}{2}\psi^*(x')(d'_0+igA_0(x'))\psi(x')-\frac{ic}{2}\psi(x')(d_0-igA_0(x'))\psi^*(x')\nonumber\\
&&-\frac{1}{2m}\left(\nabla'+ig\bs{A}(x')\right)\psi(x')\cdot\left(\nabla'-ig\bs{A}(x')\right)\psi^*(x')+f\left(\bs{\mathbb{E}}(x'),\bs{\mathbb{B}}(x')\right)
\label{3.2.16}
\end{eqnarray}
\end{widetext}
\subsection{Galilean transformations of the gauge field equations of motion}\label{sec3.3}
Using the Lagrangian density \eqref{3.2.16}, the Euler-Lagrange equations for the $A_\mu$ component of the gauge field reads
\begin{equation}
d'_\nu\left(\frac{\partial\L'}{\partial(d'_\nu A_\mu)}\right)=\frac{\partial\L'}{\partial A_\mu}\label{3.3.1}
\end{equation}
where, as usual, the summation of repeated indices is implied over the full range of their variations. 
\vskip .5cm
We first consider the $A_0$ component. Then,
\begin{eqnarray}
\frac{\partial\L'}{\partial A_0}&=&-gc\psi^*\psi\label{3.3.2}\\
\frac{\partial\L'}{\partial(d'_0A_0)}&=&0\label{3.3.3}\\
\frac{\partial\L'}{\partial(d'_iA_0)}&=&\frac{\partial f}{\partial{\bs{\mathbb{E}}_k}}\frac{\partial{\bs{\mathbb{E}}_k}}{\partial(d'_iA_0(x'))}
+\frac{\partial f}{\partial{\bs{\mathbb{B}}_k}}\frac{\partial{\bs{\mathbb{B}}_k}}{\partial(d'_iA_0(x'))}
=\frac{\partial f}{\partial{\bs{\mathbb{E}}_i}}\nonumber\\
\label{3.3.4}
\end{eqnarray}
Substituting \eqref{3.3.2}-\eqref{3.3.4} into \eqref{3.3.1} and making use of definition \eqref{2.2.12}, we obtain 
\begin{equation}
d'_i\frac{\partial f}{\partial{\bs{\mathbb{E}}_i}(x')}=-c\rho(x')\label{3.3.5}
\end{equation}
\vskip .25cm
Next, consider the Euler-Lagrange equations for a spatial component $A_i$. 
\begin{eqnarray}
\frac{\partial\L'}{\partial A_i}&=&-i\frac{g}{2m}\psi(d'_i-igA_i)\psi^*+\ i\frac{g}{2m}\psi^*(d'_i+igA_i)\psi\nonumber\\
\label{3.3.7}\\
\frac{\partial\L'}{\partial(d'_0A_i)}&=&\frac{\partial f}{\partial{\bs{\mathbb{E}}_k(x')}}\frac{\partial{{\bs{\mathbb{E}}}_k}}{\partial(d'_0A_i(x'))}
=-\frac{\partial f}{\partial{\bs{\mathbb{E}}}_i(x')}\label{3.3.8}\\
\frac{\partial\L'}{\partial(d'_jA_i)}&=&\frac{\partial f}{\partial{\bs{\mathbb{E}}_k(x')}}\frac{\partial{{\bs{\mathbb{E}}}_k}}{\partial(d'_jA_i(x'))}
+\frac{\partial f}{\partial{\bs{\mathbb{B}}_k(x')}}\frac{\partial{{\bs{\mathbb{B}}}_k}}{\partial(d'_jA_i(x'))}\nonumber\\
&=&\frac{\partial f}{\partial{\bs{\mathbb{E}}_k(x')}}(\delta_{ki}(R^{-1}\beta)_j-\delta_{kj}
(R^{-1}\beta)_i)+\frac{\partial f}{\partial{\bs{\mathbb{B}}_k(x')}}\epsilon_{kji}\nonumber\\
&=&\frac{\partial f}{\partial{\bs{\mathbb{E}}_i(x')}}(R^{-1}\beta)_j-\frac{\partial f}{\partial{\bs{\mathbb{E}}_j(x')}}(R^{-1}\beta)_i
+\epsilon_{ikj}\frac{\partial f}{\partial{\bs{\mathbb{B}}_k(x')}}\nonumber\\
\label{3.3.9}
\end{eqnarray}

Substituting \eqref{3.3.7}-\eqref{3.3.9} into \eqref{3.3.1} and making use of the definition \eqref{2.2.13}, we obtain
\begin{widetext}
\begin{equation} 
-d'_0\frac{\partial f}{\partial{\bs{\mathbb{E}}}_i(x')}
+d'_j\left((R^{-1}\beta)_j\frac{\partial f}{\partial{\bs{\mathbb{E}}_i(x')}}
-(R^{-1}\beta)_i\frac{\partial f}{\partial{\bs{\mathbb{E}}_j(x')}}\right)-\epsilon_{ijk}d'_j\frac{\partial f}{\partial{\bs{\mathbb{B}}_k(x')}}=j_i(x')
\label{3.3.10}
\end{equation}
\end{widetext}
As in \eqref{2.2.9c}, it is convenient to define the vector valued functions 
\begin{eqnarray}
\nabla_{\bs{\mathbb{E}}}f&:=&\left(\frac{\partial f}{\partial{\bs{\mathbb{E}}_1(x')}},\frac{\partial f}{\partial{\bs{\mathbb{E}}_2(x')}},
\frac{\partial f}{\partial{\bs{\mathbb{E}}_3(x')}}\right)\nonumber\\
\nabla_{\bs{\mathbb{B}}}f&:=&\left(\frac{\partial f}{\partial{\bs{\mathbb{B}}_1(x')}},\frac{\partial f}{\partial{\bs{\mathbb{B}}_2(x')}},
\frac{\partial f}{\partial{\bs{\mathbb{B}}_3(x')}}\right)\label{3.3.11}
\end{eqnarray}
and 
\begin{eqnarray}
\nabla_{\bs{E'}}f&:=&\left(\frac{\partial f}{\partial{E'_1(x)}},\frac{\partial f}{\partial{E'_2(x)}},
\frac{\partial f}{\partial{E'_3(x)}}\right)=R\nabla_{\bs{\mathbb{E}}}f\nonumber\\
\nabla_{\bs{B'}}f&:=&\left(\frac{\partial f}{\partial{B'_1(x)}},\frac{\partial f}{\partial{B'_2(x)}},
\frac{\partial f}{\partial{B'_3(x)}}\right)=R\nabla_{\bs{\mathbb{B}}}f\nonumber\\
\label{3.3.12}
\end{eqnarray}

With the definitions \eqref{3.3.11}, the equations of motion \eqref{3.3.5} and \eqref{3.3.10} can be written as vector equations:
\begin{eqnarray}
\nabla'\cdot\nabla_{\bs{\mathbb{E}}}f&=&c\rho(x')\label{3.3.13}\\
-d'_0\nabla_{\bs{\mathbb{E}}}f-R^{-1}\bs{\beta}(\nabla'\cdot\nabla_{\bs{\mathbb{E}}}f)\quad&&\nonumber\\
+(R^{-1}\bs{\beta}\cdot\nabla')\nabla_{\bs{\mathbb{E}}}f
-\nabla'\times\nabla_{\bs{\mathbb{B}}}f&=&\bs{j}(x')\nonumber\\
\label{3.3.14}
\end{eqnarray}

Multiplying these equations by the rotation matrix $R$ and using the transformation formula \eqref{3.1.10} for the differential operators and the identities 
\eqref{3.3.12}, we can rewrite \eqref{3.3.13} and \eqref{3.3.14} as equations for the transformed fields. From \eqref{3.3.13}, it follows immediately 
\begin{equation}
\nabla\cdot\nabla_{\bs{E}'}f=\rho(x')\label{3.3.15}
\end{equation}
From \eqref{3.3.14}, it follows
\begin{eqnarray}
-(d_0+\bs{\beta}\cdot\nabla)\nabla_{\bs{E}'}f-\bs{\beta}(\nabla\cdot\nabla_{\bs{E}'}f)\quad&&\nonumber\\
+(\bs{\beta}\cdot\nabla)\nabla_{\bs{E}'}f-\nabla\times\nabla_{\bs{B}'}f&=&R\bs{j}(x')\nonumber\\
-d_0\nabla_{\bs{E}'}f-\nabla\times\nabla_{\bs{B}'}f&=&R\bs{j}(x')+\bs{\beta}\rho(x')\nonumber\\
\label{3.3.16}
\end{eqnarray}
where we have used \eqref{3.3.15} in the last equality of \eqref{3.3.16}. 
Notice that  the arguments of $f$ are the transformed fields $\bs{E}'(x)$ and $\bs{B}'(x)$. The arguments of functions $\rho$ and $\bs{j}$ 
on the right hand side of \eqref{3.3.15} and \eqref{3.3.16} are $x'$, not $x$. 

The equations \eqref{3.3.15} and \eqref{3.3.16} show that the charge-current density field transform as a vector field under the $D$-representation 
\eqref{3.4} of the Galilei group. That is, if we define 
\begin{eqnarray}
\rho'(x)&=&\rho(x')\nonumber\\
\bs{j}'(x)&=&R\bs{j}(x')+\bs{\beta}\rho(x')\label{3.3.17}
\end{eqnarray}
then \eqref{3.3.15} and \eqref{3.3.16} can be written as 
\begin{widetext}
\begin{eqnarray}
\nabla\cdot\nabla_{\bs{E}'}f(\bs{E}'(x),\bs{B}'(x))&=&\rho'(x)\label{3.3.18}\\
-d_0\nabla_{\bs{E'}}f(\bs{E}'(x),\bs{B}'(x))-\nabla\times\nabla_{\bs{B'}}f(\bs{E}'(x),\bs{B}'(x))&=&\bs{j}'(x)
\label{3.3.19}
\end{eqnarray}
\end{widetext}
It is straightforward to verify that $\rho'(x)$ and $\bs{j}'(x)$ defined by \eqref{3.3.17} fulfill the continuity equation 
\begin{equation}
d_0c\rho'(x)+\nabla\cdot\bs{j}'(x)=0\label{3.3.20}
\end{equation}
Comparison of equations \eqref{3.3.18}-\eqref{3.3.20} with \eqref{2.2.8b}, \eqref{2.2.9b} and \eqref{2.2.16} establish that the full gauge field equations are Galilean invariant  for any differentiable function $f$. In particular, the choice \eqref{2.2.6d} for $f$ gives the Maxwell equations for the transformed fields $\bs{E}'$ and $\bs{B}'$: 
\begin{eqnarray}
\nabla\cdot\bs{E}'(x)&=&\rho'(x)\label{3.3.21}\\
-\frac{\partial}{\partial t}\bs{E}'(x)+\nabla\times\bs{B}'(x)&=&\bs{j}'(x)\label{3.3.22}
\end{eqnarray}
From the defining relations \eqref{3.2.7} we obtain the homogeneous equations 
\begin{eqnarray}
\nabla\cdot\bs{B'}&=&0\nonumber\\
\nabla\times\bs{E'}+\frac{\partial}{\partial t}\bs{B'}&=&0\label{3.3.23}
\end{eqnarray}
Equations \eqref{3.3.20}-\eqref{3.3.23} show that \emph{Maxwell's equations are invariant in form under Galilean transformations}. 
 
In connection with the problem of Galilean invariance of the gauge field equations, it is important to observe that both representations \eqref{3.4} and \eqref{3.6} of the group of homogeneous Galilean transformations play a crucial role. In particular,  we chose the gauge field $A_\mu$ to transform as a vector field under \eqref{3.6} because it is coupled to the differential operators $d_\mu$, and $d_\mu$ transform as a vector under \eqref{3.6} when the spacetime coordinates $x^\mu$ transform a vector under \eqref{3.4}. The transformation laws for the electric and magnetic fields follow from those of $A_\mu$ and $d_\mu$.  We also took the matter field to transform as a spinless, unitary, irreducible projective representation of the Galilei group. The transformation rules \eqref{3.3.17} for the charge-current density as a vector field under \eqref{3.4} is a consequence of the transformation properties of the matter field.  Many previous studies, such as \cite{bellac, brown}, which attempted to understand the problem as a direct limit of the relativistic transformation properties of electromagnetic fields arrived at the conclusion that full Maxwell's equations are inconsistent with Galilean relativity because such a direct limit would result in an electromagnetic  field and a charge-current density vector that transform under the same representation, either under $C$ for the `magnetic limit' or under $D$ for the `electric limit'. In these cases, Maxwell's equations would not remain Galilean invariant. In view of our result, the duality of $C$ and $D$ representations is critical the problem. The conclusions of \cite{goldin} are somewhat similar to ours, but there the Galilean invariance of Maxwell's equations is obtained  by demanding non-linear constitutive equations. Nonlinear gauge field equations can be accommodated in the theory presented here as well by choosing a suitable function $f$ in \eqref{2.2.7}. 

\section{Concluding remarks}
The motivation for this work was to examine the nature of the force that is associated with the symmetry of a non-relativistic quantum mechanical particle under local $U(1)$ gauge transformations. Not surprisingly, the Lagrangian density for a free particle, giving rise to the familiar Schr\"odinger equation $i\frac{d}{dt}\psi=-\frac{1}{2m}\nabla^2\psi$ does not remain invariant under local $U(1)$ gauge transformations, $\psi(x)\to e^{-i\lambda(x)}\psi(x)$ and 
$\psi^*(x)\to e^{i\lambda(x)}\psi^*(x)$. In order to ensure invariance, the free particle field $\left(\begin{array}{c}\psi\\ \psi^*\end{array}\right)$ must couple to a four component gauge field $A_\mu(x)$ which transforms as $A_\mu(x)\to A_\mu(x)+\frac{1}{g}d_\mu\lambda(x)$. With a suitable choice for the gauge field Lagrangian, the Euler-Lagrange equations of motion obtained from the coupled, $U(1)$ invariant  Lagrangian are precisely Maxwell's equations. This result is rather surprising because, as widely acknowledged, Maxwell's equations are relativistic equations. In fact, the exact calculation done here can be repeated starting with the Lagrangian density for the Klein-Gordon field and the result is again the very same Maxwell's equations \cite{ryder}. Nevertheless, we have shown here that they can be obtained by starting with the Schr\"odinger equation. Therefore, it appears reasonable to conclude that Maxwell's equations are the way they are not because of a particular relativity of spacetime but because of the symmetry structure of local $U(1)$ gauge transformations. 

In order to further examine this conjecture, we considered the Galilean transformation properties of the gauge invariant Schr\"odinger Lagrangian density. We took the matter field $\psi$ to furnish the usual unitary, irreducible, projective representation of the Galilei group. If we demand that the gauge field $A$ be a Galilean vector field then there are two choices: It is a vector field $A_\mu$ under the representation \eqref{3.6}, or it is a vector field $A^\mu$ under the representation \eqref{3.4}. If we take spacetime vectors $x^\mu$ transform as a Galilean vector under \eqref{3.4}, the usual choice of non-relativistic classical and quantum physics, then only the choice \eqref{3.6} for the gauge field gives rise to a Galilean invariant theory. In particular, these two choices, namely that the matter field furnishes a unitary, irreducible, projective representation of the Galilei group, and the gauge field is a Galilean vector field under the dual representation \eqref{3.6}, uniquely determine the existence of  electromagnetic fields that transform as in \eqref{3.2.7} and a current vector that transforms under \eqref{3.4}, respectively. Consequently, the resulting Maxwell's equations  are Galilean invariant. 

It remains an interesting, albeit perhaps an academic, problem to investigate the consequences of the choice that the $U(1)$ gauge field be a vector field under the representation $D$, \eqref{3.4}. In order to develop the treatment consistently, we must then require that spacetime vectors transform under \eqref{3.6}, i.e., 
\begin{eqnarray}
t'&=&t-R^{-1}\bs{v}\cdot\bs{x}+b\nonumber\\
\bs{x}'&=&R\bs{x}+\bs{a}\label{4.1}
\end{eqnarray}
The main technically demanding part of the job is to determine the unitary, irreducible, projective representations of the group defined by these transformations. Once done, we can take the matter field functions $\psi$ (which will  \emph{not} be the solutions to the Schr\"odinger equation) to transform under these representations. 
Therewith, we expect Maxwell's equations to acquire the Galilean invariant form that is dual to \eqref{3.3.20}-\eqref{3.3.23}. 

\section*{Acknowledgments}
We are grateful to Grinnell College for financial support.

\end{document}